\documentclass[12pt, fleqn]{article}
\usepackage[cp1251]{inputenc}
\usepackage{latexsym,amsfonts,amssymb}
\usepackage{graphicx}

\usepackage{amsbsy}
\usepackage{amsmath}
\usepackage{epsf}
\usepackage{cite}

\usepackage{color}

\newcommand{\plussupset}{%
  \mathrel{%
    \ooalign{%
      $\supset$\cr
      \hidewidth\kern-0.1em\raisebox{0.09ex}{$\scriptstyle+$}\hidewidth\cr
    }%
  }%
}

\newtheorem{theorem}{Theorem}
\newtheorem{definition}{Definition}
\newtheorem{remark}{Remark}
\newcommand{\bt}{\begin{theo}}
\newcommand{\et}{\end{theo}}
\newcommand{\bd}{\begin{displaymath}}
\newcommand{\ed}{\end{displaymath}}

\newcommand{\be} {\begin{equation}}
\newcommand{\ee} {\end{equation}}
\newcommand{\ba} {\begin{array}{l}}
\newcommand{\ea} {\end{array}}
\newcommand{\bea}{\begin{eqnarray}}
\newcommand{\eea} {\end{eqnarray}}

\newcommand{\p} {\partial}

\begin{document}

\begin{center}
 {\Large \bf
Symmetries and exact solutions of a reaction-diffusion system
arising in population dynamics}

\medskip

{\bf Roman Cherniha $^1,^2$, Philip Broadbridge $^3$,\\  Vasyl'
Davydovych $^4$ and Ian Marquette $^5$ }

$^1$ Department of Mathematics, National University of Kyiv-Mohyla
Academy,
  2 Skovoroda Street,  04070 Kyiv, Ukraine

$^2$School of Mathematical Sciences, University of Nottingham,
  University Park, Nottingham NG7 2RD, UK

$^3$Department of Mathematical and Physical Sciences, La Trobe
University, Bundoora VIC 3086, Australia

$^4$Institute of Mathematics,  NAS  of Ukraine,
 3, Tereshchenkivs'ka Street, Kyiv 01004, Ukraine

$^5$Department of Mathematical and Physical Sciences, La Trobe
University, Bendigo VIC 3552, Australia

\emph{Email: r.m.cherniha@gmail.com, \ p.broadbridge@latrobe.edu.au,
\ davydovych@imath.kiev.ua, \ i.marquette@latrobe.edu.au}

\end{center}

\begin{abstract}
A system
    of two cubic reaction-diffusion equations for two independent gene frequencies
    arising in population dynamics is studied. Depending on values of coefficients,
    all possible Lie and
    $Q$-conditional (non-classical) symmetries are identified. A wide
    range of new exact solutions is constructed, including those expressible in terms of  a Lambert
    function and not obtainable by Lie symmetries.
    An example of a new
real-world application of the system is discussed as well.

\end{abstract}



\section{Introduction}\label{sec-1}

Systems of reaction-diffusion (RD) equations are ubiquitous in
continuum models of chemical reactions, population ecology,
 population genetics, epidemiology and oncology (see well-known books
  \cite{aris-75I,britton,ch-dav-book, fife-79,ku-na-ei-16, mur2, mur2003, okubo} and
  hundreds of papers cited therein). Compared to the theory of scalar densities, less is known about the behaviour of two- and multi-component
 systems. For example, the densities of two interacting chemical reagents may vary over space and time. In ecology there may be two species
  that display competition, mutualism or predator-prey dynamics. There may be three possible alleles that can occupy a locus in the genetic DNA code of a diploid species. In that case there will be five possible genotypes but
  as shown by Bradshaw-Hajek \textit{et al.} \cite{broadbridge08}, when all phenotypes have the same level of mobility, there is a closed system of two cubic RD equations for two independent gene frequencies within the gene pool.\\

There have been a number of numerical simulations of RD systems.
Exact solutions and reduction of order have been relatively scarce.
The focus here is on the symmetries and reductions of two-component
systems.
 From \cite{broadbridge08} after
the notation change \[ p_1 \rightarrow u, \quad p_2 \rightarrow v,
\] the allele frequencies
 $u$ and $v$ in one spatial dimension
satisfy
\begin{equation}\label{1-1}\ba
u_t=d_1 u_{xx}  +\frac 2\rho \rho_x u_x+\Phi(u,v), \medskip \\
v_t=d_2 v_{xx}  +\frac 2\rho \rho_x v_x+\Psi(u,v),
 \ea\end{equation}
 as well as a trivial balance equation for $w$ that is $w=1-u-v$.
The analysis in \cite{broadbridge08} assumed $d_1=d_2$, leading to
source terms  that can be cubic or quadratic, depending on the size
of genotype fitness coefficients $\gamma_{ij}$, with $i,j=1,2,3$
denoting the presence of paired alleles whose density adds to
$u,v,w$.
\begin{equation}\label{PhiPsi}\ba
\Phi(u,v)=u(\gamma_{13}-\gamma_{33})+u^2(\gamma_{11}-3\gamma_{13}+2\gamma_{33})+u^3(-\gamma_{11}+2\gamma_{13}-\gamma_{33})\\
\hskip2cm+uv(\gamma_{12}-\gamma_{13}-2\gamma_{23}+2\gamma_{33})+u^2v(-2\gamma_{12}+2\gamma_{13}+2\gamma_{23}-2\gamma_{33})\\
\hskip3cm +uv^2(-\gamma_{22}+2\gamma_{23}-\gamma_{33}),\medskip \\
\Psi(u,v)=v(\gamma_{23}-\gamma_{33})+v^2(\gamma_{22}-3\gamma_{23}+2\gamma_{33}) +v^3(-\gamma_{22}+2\gamma_{23}-\gamma_{33})\\
\hskip2cm+uv(\gamma_{12}-\gamma_{23}-2\gamma_{13}+2\gamma_{33}) + uv^2(-2\gamma_{12}+2\gamma_{23}+2\gamma_{13}-2\gamma_{33})\\
 \hskip3cm+  u^2v(-\gamma_{11}+2\gamma_{13}-\gamma_{33}). \ea\end{equation}

Whereas genotype-dependent mobilities do not lead exactly to
(\ref{PhiPsi}), there are advantages in allowing $d_1\ne d_2$.
 Firstly, this generality may allow a richer symmetry structure of the equations while the solutions are not highly sensitive
 to the ratio $d_1/d_2$ compared to the effect of ratios of the fitness coefficients that modify the source terms.
  Secondly, in the case that $u$ is the density of a dominant new gene with improved fitness,
  the expression for $\Phi(u,v)$ will be correct while the expression for $\Psi(u,v)$ will be approximately correct
  when $(u(x,t),v(x,t))$ evolves towards $(1,0)$.

  In this work, we apply  symmetry-based methods \cite{bluman2010,ch-se-pl-book, olv-93} for a nonlinear
  RD system that naturally follows from
  (\ref{1-1})--(\ref{PhiPsi}).
It should be stressed that  nowadays symmetry-based methods  are
widely applied for analysis of nonlinear evolution systems arising
in mathematical modelling of real-world processes, in particular, in
physics \cite{ch-ki-24, ch-ki-25, oliveri-et-al-25,
lo-dimas-bo-2023,soph-24}, ecology
\cite{ch-da-AAM-23,broadbridge23,torrisi-23}, biomedicine
\cite{ch-dav-book,ch-da-EJAM-22,naz-24,rosa-23,torrisi-21}.

The paper is organised as follows. Section~\ref{sec-2}, we present
the main body of results which consists in the classification of Lie
and conditional symmetries. The non-trivial cases are detailed in
Theorem~\ref{th-1}.
 In Section~\ref{sec-3}, we provide the exact solutions of the system of nonlinear PDEs.
  This leads to new reductions,
 in particular to systems of ODEs that are integrable and which admit exact solutions in terms
 of the Lambert $W$ function.  Notably, we were able to construct both
 Lie's solutions and non-Lie solutions (i.e. the exact solutions
 that cannot be derived via Lie symmetries).
In Section~\ref{sec-4}, the main theoretical result,
Theorem~\ref{th-1}, is proved. It is shown that
a very complicated system consisting  of two overdetermined
subsystems (each involves  12 differential equations) are completely
integrable and, as a result, Theorem~\ref{th-1} is obtained.
Finally, we briefly discuss the results obtained and present some
conclusions  in the last section.

\section{Lie and conditional symmetries: main results}\label{sec-2}

The case of most interest is that of  Mendelian inheritance wherein
the third gene   dominates everything else, and the second gene
dominates the first, therefore \[\ba
\gamma_{33}=\gamma_{32}=\gamma_{31}=g_{33}, \\
 \gamma_{22}=\gamma_{21}=g_{22},\quad \gamma_{11}=g_{11} ,\quad
A=g_{33}-g_{11} \geq 0 ,\quad B=g_{33}-g_{22}\geq 0. \ea\] Under the
above restrictions, the RD system (\ref{1-1}) takes the form
\begin{equation}\label{2-2*}\ba  u_t =  u_{xx} - A u^2 (1-u) - B u v +2 B u^2 v + B u
v^2,
\\
 v_t =  v_{xx} - B v^2 (1-v) - B u v + 2 B u v^2 + A u^2 v.
\ea\end{equation}

In contrast to \cite{broadbridge08},  we do not assume that the
genes have the same diffusivities, therefore system (\ref{2-2*})
takes more general form
\begin{equation}\label{2-2}\ba  u_t = d_1 u_{xx} - A_1u^2 (1-u) - B_1 u v +2 B_1 u^2 v
+ B_1 u v^2,
\\
 v_t = d_2 v_{xx} - B_2 v^2 (1-v) - B_2 u v + 2 B_2 u v^2 + A_2 u^2v, \ea\end{equation}
 where $d_i>0$,  while  $A_i$ and $B_i, \ i=1,2$ are non-negative constants
with $(A_i,B_i)\ne (0,0)$.

 Lie  symmetries of  the RD system (\ref{2-2}) can be identified
 from papers \cite{2-ch-king1} and \cite{2-ch-king2}, in which a
 complete   classification of Lie symmetries of the
 general class of two-component RD systems is derived.
So, (\ref{2-2}) with $B_1=B_2=0$, $d_1 \neq d_2$  admits a
three-dimensional Lie algebra with the basic operators (see Table 4
in \cite{2-ch-king1})
\[ \langle\partial_t, \partial_x, v \partial_v \rangle, \]
and that with $B_1=B_2=0$, $d_1=d_2$ admits a four-dimensional Lie
algebra with the basic operators (see Table 3 in \cite{2-ch-king2})
\begin{equation}\label{2-2a} \langle \partial_t,\partial_x,v \partial_v,
(u-1) \partial_v\rangle.
\end{equation}
These operators respectively generate translations in time and
space, scaling of $v$ and linear mixing of components, $\bar
v=v+\epsilon(u-1).$  In the case $B_1^2+B_2^2\not=0$, system
(\ref{2-2}) admits the non-trivial Lie algebra only if diffusivities
are the same, therefore one may reduce those to $d_1=d_2=1$.  Having
done this  and setting
\[ A_1=A_2=B_1=B_2 \equiv A, \]
one obtains the system
\begin{equation}\label{2-2**}\ba  u_t =  u_{xx} - A u^2 (1-u) - A u v +2A u^2 v + A u
v^2,
\\
 v_t =  v_{xx} - A v^2 (1-v) - A u v + 2 A u v^2 + A u^2 v.
\ea\end{equation}

In the context of successive dominance of three alleles, the special
case $A=B$ implies $g_{22}=g_{33}$, so the second and third alleles
 have equal phenotype advantage. In other applications $A$ may be negative.

 It may be proved by direct calculations that the above system
 admits the following  highly non-trivial Lie algebra of invariance
\[\langle \partial_t,\partial_x,u\left(\partial_u-\partial_v\right),
v\left(\partial_u-\partial_v\right)\rangle.\]

Applying  the substitution
\begin{equation}\label{2-33} t \rightarrow \frac{1}{|A|}\,t, \
x\rightarrow\sqrt{\frac{1}{|A|}}\,x, \ u\rightarrow \frac{u+v}{2}, \
v\rightarrow \frac{u-v}{2},\end{equation}system (\ref{2-2**}) is
transformed to the RD system
\begin{equation}\label{2-31}\ba  u_t = u_{xx} - \delta u^2(1-u), \\
 v_t = v_{xx} - \delta uv(1-u), \ea\end{equation}
where $\delta=sign(A)$.

 A possible interpretation of the above
RD system can be as follows. For example with $\delta<0$, this may
represent commensalism. In that application, $u$ represents the
population density of a predator species, while $v$ represents the
population density of a benefiting commensal species with similar
mobility. The classic example is that of tigers and
carrion-consuming jackals in the past when the former were
plentiful. The $u^2$ factor denotes a weak Allee effect as prey can
more easily find an escape path when there are few predators in the
vicinity. When the predators reach environmental carrying capacity,
scaled here to $u=1$, they cannot afford to discard morsels of food
for the carrion consumers. The latter then relocate in search of a
more reliable food supply before they reproduce.

 System (\ref{2-31}) is
again a particular case of that from Table 3 \cite{2-ch-king2} and
admits the Lie algebra with the basic operators
\begin{equation}\label{2-32} \langle \partial_t,\partial_x,u\partial_v, v\partial_v\rangle,\end{equation}
which is, of course, equivalent to (\ref{2-2a}).

There are no other special cases when system (\ref{2-2})
 with $B_1^2+B_2^2\not=0$ admits a non-trivial Lie symmetry generated by the operators $\partial_t$ and
$\partial_x$.

Now we are looking for $Q$-conditional (non-classical)  symmetries
\cite{bluman2010,ch-dav-book} of the RD system~(\ref{2-2}).


The most general form of a $Q$-conditional symmetry has a generating
operator
\begin{equation}\label{2-3}  Q= \xi^0(t,x,u,v)\partial_t + \xi(t,x,u,v) \partial_x +
\eta^1(t,x,u,v) \partial_u + \eta^2(t,x,u,v) \partial_v,
\end{equation}
where the functions $\xi^0, \ \xi, \ \eta^1$ and $\eta^2$ should
satisfy the relevant criteria of $Q$-conditional invariance. The
motivation here is that symmetry reductions and invariant solutions
of (\ref{2-2}) may be constructed even though the operator $Q$ does
not leave that system invariant in Lie's sense. It is well-known
that one should examine two essentially different cases
$\xi^0\not=0$ and $\xi^0=0$ in order to find such symmetries.

\begin{theorem}\label{th-1} The system RD (\ref{2-2}) admits a $Q$-conditional
symmetry of the form (\ref{2-3})  with $\xi^0\not=0$ only in the
case $d_1\not=d_2, \ B_1=B_2=0$ and $A_1d_2=A_2d_1$, i.e.
 \begin{equation}\label{2-4}\ba u_t
= d_1 ( u_{xx} - A u^2 (1-u) ), \\ v_t = d_2 ( v_{xx} + A u^2 v ).
\ea \end{equation}
  The most general form of the
$Q$-conditional symmetry ($\xi^0\not=0$) of system (\ref{2-4}) is
given by the formula
\begin{equation}\label{2-5}
 Q= \partial_t + \gamma \partial_x + \Big( \alpha \exp( \sigma(x + d_2
\sigma t))(u-1) + \beta v \Big) \partial_v, \end{equation} where
$A$, $\alpha$, $\beta$, $\gamma$ are arbitrary constants and $\sigma
= \gamma \frac{d_2 -d_1}{2 d_1 d_2}. $
\end{theorem}

It should be noted that the above symmetry with $\gamma=0$
simplifies to the form
\[Q = \partial_t + (\alpha(u-1) + \beta v ) \partial_v. \]
The latter is still a $Q$-conditional symmetry (not a Lie symmetry!)
of (\ref{2-4}).

Operator (\ref{2-5}) with  $d_1=d_2$, i.e.
\begin{equation}\label{2-5*} \partial_t + \gamma \partial_x + (\alpha (u-1) + \beta v )
\partial_v
\end{equation}
is a Lie symmetry because that is nothing else but  a linear
combination of the Lie symmetries from (\ref{2-2a}).

In the case of $\xi^0=0$, all possible  $Q$-conditional symmetries
of the RD system (\ref{2-2}) cannot  be identified  in the general
case because the problem is reducible to solving the system in
question (for evolution scalar equations it was proved in
\cite{zhdanov}, see also \cite{cherniha21b} for evolution systems).
As a result, only particular $Q$-conditional symmetries can be
constructed under the correctly-specified restrictions in the
so-called no-go case.

An essential progress in analysis of the no-go case can be done
applying  the notion of {\it $Q$-conditional symmetry of the first
type}\cite{ch-2010} (see also \cite{cherniha21b}). Symmetries of
this type will be constructed for the RD system (\ref{2-2}) in a
forthcoming work.

\subsection{An example:   settling in a new mining town}

 Here we present another  example for a realistic interpretation of the
RD system (\ref{2-4}) with $A_1=Ad_1$, which is related to  human
populations and  non-renewable resources. So, we consider the system
\begin{equation}\label{5-1**} \ba
u_t=d_1u_{xx}-A_1u^2(1-u), \\
v_t=d_2v_{xx}+A_1\frac{d_2}{d_1}u^2v.\ea \end{equation}

For example with $A_1<0$,  $u$ may be a human population density of
settlers in the region of a newly established mine in a region
 of high concentration $v$ of a non-renewable resource, such as  rare metals in beach sand.  The human population carrying capacity
  is scaled to unity. The mineral migrates diffusively by erosion and is harvested by mining. The $u^2$ factors denote a weak Allee effect,
  as new settlers require a number of cooperating members to set up a new mine as well as new accommodation. \\
The first equation in (\ref{5-1**}) has the cubic reaction term of a
Huxley equation. The Huxley RD equation for $u$ is decoupled from
$v$. The non-classical symmetries of that scalar equation are fully
known \cite{a-h-b,cla-mans-94}, and explicit non-classical invariant
 solutions were used in population genetics \cite{h-b}.
 Under the special restriction $A_2=A_1d_2/d_1$, invariant solutions under
 the $(u,v)$ system's additional Q-conditional symmetry,
 have the  travelling  wave form $u=f(x-\gamma t)$.

  The steady state $u(x)$ satisfies a second-order autonomous ODE with no first-order term.
  Just as for Newton's force equation in one dimension, it has an energy integral, which in this case is
\[ \mathcal E=d_1\frac{u_x^2}{2}-A_1\int_0^u \bar u^2(1-\bar u)d\bar u.\]
For a symmetric solution with identical boundary conditions at
$x=\pm L$, $u_x(0,t)=0$ and
\[ \mathcal E=A_1\left(\frac{u_*^4}{4}-\frac{u_*^3}{3}\right)>0,\] where $u_*$ is the central
maximum value attained at $x=0$. Hence, the symmetric solution in
inverse form, is expressible as an elliptic integral
\[x=\pm \sqrt{\frac{d_1}{2}}
\int_u^{u_*}\frac{d\bar u}{\sqrt{\mathcal E+A_1(\frac{\bar
u^3}{3}-\frac{\bar u^4}{4})}}.\] The minimum length of a domain to
support a non-uniform positive steady state \cite{BHH}
 is approximately $6.42\times \sqrt{-d_1/A_1}$.
This domain enlarges as the central maximum value $u_*$ increases
above the critical
 value $0.58$. For example with
the central population density $u_*=0.998$ near carrying capacity
\cite{BHH}, the non-zero population density extends to the boundary
of the compact support at $x\approx\pm 2.35\times3.21\sqrt{-d_1/A_1}$.\\
 Consider a settlement population growth coefficient of $-A_1=2$ yr$^{-1}$, and diffusion coefficient $d_1=$ 2 km$^2$ yr$^{-1}$.
  For a spatially uniform population, $u(x,t)$ satisfies the separable ODE $ u_t+A_1u^2(1-u)=0 $, with solution
 \[-A_1t=\log\frac{u(1-u_0)}{u_0(1-u)}+\frac{1}{u_0}-\frac{1}{u}.\]
  The time taken to increase the settled population density $u$ from $0.2$ to $0.8$ will be 6.5 years.
 At the minimum size of a compactly supported steady state, the town limits are at $x=\pm 3.2$ km.
 If the central density reaches a high value of 0.998, the town would now be a city extending to $x=\pm 7.5$ km.
  This might be the case if the mine has a long expected longevity such as $1/A_2=40$ yr.

\section{Exact solutions of the RD system (\ref{2-2**})}\label{sec-3}

 We point out from the very beginning that a
variety of exact solutions of the RD system (\ref{2-2**}) can be
constructed using the steady-state points $(1,v_0)$ and $(0,v_0), \
v_0 \in \textbf{R}$ of system (\ref{2-31}). In fact, setting $u=1$
in (\ref{2-31}), the following solution of the RD system
(\ref{2-2**}): \[u(t,x)=\frac{1+g(t,x)}{2}, \
v(t,x)=\frac{1-g(t,x)}{2}\] is easily derived.  Here  the function
$g(t,x)$ is an arbitrary solution of the standard classical heat
equation $g_t=g_{xx}.$ Similarly, $u=0$  leads to
\[u(t,x)=\frac{g(t,x)}{2}, \
v(t,x)=-\frac{g(t,x)}{2}.\] The latter  is questionable for
interpretation because one of the components is negative.

Another peculiarity of system (\ref{2-31}) follows from its
structure and allows to use known exact solutions of the first
equation. In fact, exact solutions  of this equation for the first
time were   identified in \cite{a-h-b, cla-mans-94} (see also
Section 4.3.2 in  \cite{ch-se-pl-book}). Notably this equation with
$\delta=-1$  is referred to as the Huxley equation. So, having a
known solution $u_0(t,x)$, one obtains the linear heat equation with
the
source/sink $U(t,x)=\delta u_0(t,x)(1-u_0(t,x))$:
\[
 v_t = v_{xx} - U(t,x) v. \]
 Thus, depending on the form $U(t,x)$  and given boundary and/or
 initial conditions classical methods for solving linear PDEs can be
 applied.

\subsection{Lie's solutions}

 Let us apply the Lie symmetry (\ref{2-32}) to
construct exact solutions of the RD system (\ref{2-31}). The most
general linear combination of the Lie symmetries (\ref{2-32}) has
the form \be\label{3-1}
X=\alpha_1\p_t+\alpha_2\p_x+\left(\beta_1u+\beta_2v\right)\p_v,\ee
where $\alpha_1, \ \alpha_2, \ \beta_1$ and $\beta_2$ are arbitrary
constants, $\alpha_1^2+\alpha_2^2\neq0$. Depending on parameters
$\alpha_i$ and $\beta_2$ one can obtain essentially different
ans\"atze for reduction of system (\ref{2-31}) to ODE  systems.

\textbf{\emph{In the case $\alpha_1\neq0$}} operator (\ref{3-1})
takes the form
\[X=\p_t+\alpha\p_x+\left(\beta_1u+\beta_2v\right)\p_v\]
and produces two inequivalent ans\"atze and the relevant ODE systems
\begin{equation}\label{3-2}\ba
u(t,x)=\varphi(\omega), \ \omega=x-\alpha t,\\
v(t,x)=e^{\beta_2t}\psi(\omega)-\frac{\beta_1}{\beta_2}\,\varphi(\omega),\ea\end{equation}
\begin{equation}\label{3-3}\ba
\varphi''+\alpha\varphi'+\delta\varphi^2\left(\varphi-1\right)=0,\\
\psi''+\alpha\psi'-\psi\left(\beta_2+\delta\varphi-\delta\varphi^2\right)=0,\ea\end{equation}
in the subcase $\beta_2\neq0,$ and
\begin{equation}\label{3-4}\ba
u(t,x)=\varphi(\omega), \ \omega=x-\alpha t,\\
v(t,x)=\psi(\omega)+\beta_1t\varphi(\omega),\ea\end{equation}
\begin{equation}\label{3-5}\ba
\varphi''+\alpha\varphi'+\delta\varphi^2\left(\varphi-1\right)=0,\\
\psi''+\alpha\psi'-\varphi\left(\beta_1+\delta\psi-\delta\varphi\psi\right)=0,\ea\end{equation}
in the subcase $\beta_2=0.$

 The structure of exact solutions of the above ODE
systems mainly depends on the parameters $\alpha$ and $\delta$.
\medskip\\
 The general solution of the first equation of the ODE system
(\ref{3-3}) with $\alpha=0$ can be presented in an implicit form in
terms of an elliptic integral,
\[\omega+C_2=\pm\int\frac{1}{\sqrt{-\frac{\delta}{2}\,\varphi^4+\frac{2\delta}{3}\,\varphi^3+C_1}}\,d\varphi.\]
Setting $C_1=0$ and $C_2=0$  (this constant reflects the invariance
of the system in question with respect to space translations), the
above integral can be expressed in terms of elementary functions:
$-\frac{\sqrt{3\varphi^3(4-3\varphi)}}{\sqrt{2}\varphi^2}$ if
$\delta=1$ and
$\frac{\sqrt{3\varphi^3(-4+3\varphi)}}{\sqrt{2}\varphi^2}$ if
$\delta=-1$. As a result, one obtains  the function $\varphi$ in the
explicit form $\varphi=\frac{12}{2\delta \omega^2+9}$.

 Thus, using the ans\"atze  (\ref{3-2})  and
(\ref{3-4}),  and transformations (\ref{2-33}), the following exact
solutions  of the RD system (\ref{2-2**}) are constructed:
\[\ba
u(t,x)=\frac{6(1+\beta_1)}{2Ax^2+9}+e^{\beta_2t}\psi(x),\medskip\\v(t,x)=\frac{6(1-\beta_1)}{2Ax^2+9}-e^{\beta_2t}\psi(x),
\ea\] and \[\ba
u(t,x)=\frac{6(1+\beta_1\,t)}{2Ax^2+9}+\psi(x),\medskip\\v(t,x)=\frac{6(1-\beta_1\,t)}{2Ax^2+9}-\psi(x),
\ea\]
 where the function $\psi(x)$ is an arbitrary solution of the
linear ODEs
\[\psi''+\left(\frac{12A\left(3-2Ax^2\right)}{\left(2Ax^2+9\right)^2}-\gamma\right)\psi=0\] and
\[\psi''+\frac{12A\left(3-2Ax^2\right)}{\left(2Ax^2+9\right)^2}\,\psi-\frac{6\beta_1}{2Ax^2+9}=0,\]
respectively.

\begin{remark}
The constant solutions $\varphi=1$ and $\varphi=0$ of the first
equation of (\ref{3-3})  and (\ref{3-5}) lead to  very particular
cases of those obtained via  steady-state solutions  $u=1$ and $u=0$
(see above).
\end{remark}

In the case $\delta=-1$, the following exact solution of the first
equation of  (\ref{3-3}) can be found  in the explicit  form
\be\label{3-32}
\varphi(\omega)=\frac{1}{1+C_1\,e^{\alpha\,\omega}},\ee
where $\alpha=\pm\frac{1}{\sqrt{2}}.$ Using the obtained  function
$\varphi$ and integrating the second equations of systems
(\ref{3-3}) and (\ref{3-5}), one easily  finds  the function $\psi$.
Here we consider case $\alpha=\frac{1}{\sqrt{2}}$ and solve the
second equation of system (\ref{3-5}) \be\label{3-34}\psi(
\omega)=C_2+\frac{2\beta_1-C_2}{C_1}\,e^{-\frac{\omega}{\sqrt{2}}}+\frac{C_3+\sqrt{2}(C_2-\beta_1)\omega}{1+C_1\,e^{\frac{\omega}{\sqrt{2}}}}.\ee

Using formulae (\ref{3-4}), (\ref{3-32}), (\ref{3-34}) and
(\ref{2-33}), we derive the exact solution of the nonlinear RD
system (\ref{2-2**}) with $A<0$ in the following form
\[\ba
u(t,x)=\frac{1}{2+2C_1e^{\kappa}}\left(1+\sqrt{-2A}\,\beta_1\,x+2C_1\beta_1e^{\kappa}+C_2\left(e^{-\kappa}-C_1^2e^{\kappa}-2C_1\kappa\right)+C_3\right),\medskip \\
v(t,x)=\frac{1}{2+2C_1e^{\kappa}}\left(1-
\sqrt{-2A}\,\beta_1\,x-2C_1\beta_1e^{\kappa}-C_2\left(e^{-\kappa}-C_1^2e^{\kappa}-2C_1\kappa\right)-C_3\right),\ea\]
where $C_1, \ C_2, \ C_3$ and $\beta_1$ are arbitrary constants,
while $\kappa=\frac{At+\sqrt{-2A}\,x}{2}.$

\textbf{ \emph{In the case $\alpha_1=0, \ \alpha_2=1$}} operator
(\ref{3-1}) also produces two inequivalent ans\"atze and the
relevant ODE systems
\begin{equation}\label{3-6}\ba
u(t,x)=\varphi(t),\\
v(t,x)=e^{\beta_2x}\psi(t)-\frac{\beta_1}{\beta_2}\,\varphi(t),\ea\end{equation}
\begin{equation}\label{3-7}\ba
\varphi'+\delta\varphi^2\left(1-\varphi\right)=0,\\
\psi'-\psi\left(\beta_2^2-\delta\varphi+\delta\varphi^2\right)=0,\ea\end{equation}
in the subcase $\beta_2\neq0,$ and
\begin{equation}\label{3-8}\ba
u(t,x)=\varphi(t), \\
v(t,x)=\psi(t)+\beta_1x\varphi(t),\ea\end{equation}
\begin{equation}\label{3-9}\ba
\varphi'+\delta\varphi^2\left(1-\varphi\right)=0,\\
\psi'+\delta\varphi\psi\left(1-\varphi\right)=0,\ea\end{equation} in
the subcase $\beta_2=0.$

The equation on the function $\varphi$ in the above ODE systems is
integrable via the Lambert function and has the solution
\begin{equation}\label{3-14}
\varphi(t)=\frac{1}{\mathrm{LambertW}\left(C_1e^{\,
\delta\,t}\right)+1}.
\end{equation}
Substituting (\ref{3-14}) into the second equations of the ODE
systems (\ref{3-7}) and (\ref{3-9}) and integrating the equations
obtained, we find the functions
\begin{equation}\label{3-15}
\psi(t)=\frac{C_2\,e^{\beta_2^2\,t}}{\mathrm{LambertW}\left(C_1e^{\,
\delta\,t}\right)+1},
\end{equation} and
\begin{equation}\label{3-16}
\psi(t)=\frac{C_2}{\mathrm{LambertW}\left(C_1e^{\,
\delta\,t}\right)+1},
\end{equation}
respectively.

By using the obtained functions $\varphi$  from (\ref{3-14}) and
$\psi$ from (\ref{3-15})--(\ref{3-16}), ans\"atze (\ref{3-6}) and
(\ref{3-8}), and the substitution (\ref{2-33}), we derive the
multiparameter families of exact solutions for the nonlinear RD
system (\ref{2-2**}) in the following forms
\begin{equation}\label{3-35}\ba u(t,x)=\frac{1-\beta_1+C_2\exp\left(\beta_2x+\beta_2^2\,t\right)}{2\mathrm{LambertW}\left(C_1e^{A\,t}\right)+2},\medskip\\
v(t,x)=\frac{1+\beta_1-C_2\exp\left(\beta_2x+\beta_2^2\,t\right)}{2\mathrm{LambertW}\left(C_1e^{A\,t}\right)+2},
\ea\end{equation} and
\begin{equation}\label{3-36}\ba u(t,x)=\frac{1+\beta_1x+C_2}{2\mathrm{LambertW}\left(C_1e^{A\,t}\right)+2},\medskip\\
v(t,x)=\frac{1-\beta_1x-C_2}{2\mathrm{LambertW}\left(C_1e^{A\,t}\right)+2},
\ea\end{equation} where $C_1, \ C_2, \ \beta_1$ and $\beta_2\neq0$
are arbitrary constants. In order to avoid singularities, $C_1$ may
be chosen to be positive and the Lambert function may be restricted
to its principal branch. Having $C_1>0$, one observes that $u+v < 1$
independently of other parameters  arising in
(\ref{3-35})--(\ref{3-36}).

In conclusion of this part, we point out the following observation.
 The general technique  for constructing exact solution using Lie symmetries is well-known from the theoretical
  point of view. There are two different approaches described, e.g., in Section 1.3 \cite{ch-se-pl-book}.
  Here, the method based on the analysis of the most general invariant surface condition
  \begin{equation}\nonumber\ba \alpha_1u_t+\alpha_2u_x=0,\\
  \alpha_1v_t+\alpha_2v_x=\beta_1u+\beta_2v,
\ea\end{equation} corresponding to operator (\ref{3-1}),  was used.
   There is another method
    based on construction of optimal systems of inequivalent (non-conjugated) subalgebras of the given
   Lie algebra of symmetries, which is widely used as well (although very often without proper citations).
 All  optimal systems  of four-dimensional  algebras were described in a
seminal work  \cite{Pat-Wint}. The Lie algebra (\ref{2-32})
presented in Table~2 \cite{Pat-Wint} (see algebra $A_2\oplus2A_1$
therein). However, the relevant subalgebras have inconvenient form
for applications, therefore we used a simpler approach. Notably, we
derived four inequivalent cases that coincide with the number of
(non-conjugated) subalgebras in \cite{Pat-Wint}.

\subsection{Non-Lie solutions }

Now we turn to the $Q$-conditional symmetry (\ref{2-5}) in order to
construct exact solutions of the RD system (\ref{2-4}) that are not
obtainable by using Lie symmetries. By a standard procedure one
constructs the ansatz corresponding to (\ref{2-5}):
\begin{equation}\ba\label{3-52}
u(t,x)=\varphi(\omega), \ \omega=x-\gamma t, \medskip\\
v(t,x)=\frac{4\alpha d_1^2 d_2\left(1-\varphi(\omega)\right)}{4d_1^2
d_2 \beta + (d_1^2 - d_2^2) \gamma^2}\exp( \sigma(x + d_2 \sigma
t)+\psi(\omega)e^{\beta t}, \ea \end{equation} provided  $4d_1^2 d_2
\beta + (d_1^2 - d_2^2) \gamma^2\neq0$. In the case
$\beta=\frac{(d_2^2 - d_1^2) \gamma^2}{4d_1^2 d_2}$, one readily
obtains the ansatz
\begin{equation}\ba\label{3-53}
u(t,x)=\varphi(\omega), \ \omega=x-\gamma t, \medskip\\
v(t,x)=\alpha \left(1-\varphi(\omega)\right)\,t\exp( \sigma(x + d_2
\sigma t)+\psi(\omega)\exp(\beta t) \ea \end{equation} (we remind
the reader that $\sigma = \gamma \frac{d_2 -d_1}{2 d_1 d_2}$).

 The first component $u=\varphi(\omega)$ in as\"atze (\ref{3-52}) and (\ref{3-53}), reduces  the first equation of system
(\ref{2-4}) to the nonlinear second-order  ODE
\begin{equation}\label{3-54}
\varphi''+\frac{\gamma}{d_1}\varphi'+A_1\varphi^2\left(\varphi-1\right)=0,\end{equation}
while the second component, $v(t,x)$, yields the following
equations\,:
\[
\psi''+\frac{\gamma}{d_2}\psi'+\left(A_1\varphi^2-\frac{\beta}{d_2}\right)\psi=0,\]
and
\begin{equation}\label{3-56}
\psi''+\frac{\gamma}{d_2}\psi'+(A_1\varphi^2+
\beta)\psi+\frac{\alpha}{d_2}\exp(\sigma\,\omega)(\varphi-1)=0,\end{equation}
respectively.

To the best of our knowledge, non-constant exact solutions of
(\ref{3-54}) with $A_1>0$  are unknown. In the case $A_1<0$, the
following known exact solution  can be identified:
\begin{equation}\label{3-57}
\varphi(\omega)=
\frac{1}{1+C_1\,\exp\left(\pm\sqrt{\frac{-A_1}{2}}\,\omega\right)},
\quad \gamma= \pm d_1\sqrt{\frac{-A_1}{2}}.
\end{equation}

Substituting   (\ref{3-57}) into equation (\ref{3-56}), one readily
constructs the general solution  in terms of hyper-geometric
functions using computer algebra packages, for example,  MAPLE.  In
order to avoid cumbersome formulae, we present here a particular
solution using the above function $\varphi(\omega)$ with the sign
$+$. In this case, the exact solution of equation (\ref{3-56}) is
\begin{equation}\label{3-58}\ba
\psi(\omega)=\frac{\alpha}{3d_2\nu^2C_1^2}\frac{\exp\left(-\frac{d_1+3d_2}{2d_2}\nu\,\omega\right)}{1+C_1\,e^{\nu\,\omega}}
\Big[1+C_1e^{\nu\,\omega}-2C_1^2e^{2\nu\,\omega}+C_1^3\left(\nu\,\omega-3\right)e^{3\nu\,\omega}\medskip\\
\hskip3cm
+2\left(1+C_1\,e^{\nu\,\omega}\right)^3\ln\left(1+C_1\,e^{\nu\,\omega}\right)\Big]
,\ea\end{equation} where $\nu=\sqrt{\frac{-A_1}{2}}.$

Thus, taking into account  (\ref{3-53}),  (\ref{3-57}), and
(\ref{3-58}), we  construct the exact solution of the RD system
(\ref{2-4})
\[\ba u(t,x)=\frac{1}{1+C_1\,e^{\nu\left(x-d_1\nu\,t\right)}},
\medskip\\
v(t,x)=\frac{\alpha
C_1\,t\exp\big[\left(\sigma+\nu\right)\,x+d_2\left(\sigma^2+2\nu\sigma-\nu^2\right)t\big]}{1+C_1\,e^{\nu\left(x-d_1\nu\,t\right)}}+
\frac{\alpha\exp\big[\left(\sigma-2\nu\right)\,x+d_2\left(\sigma^2-4\nu\sigma+2\nu^2\right)\,t\big]}
{3d_2\nu^2C_1^2\left(1+C_1\,e^{\nu\left(x-d_1\nu\,t\right)}\right)}
\Big[1-2C_1^2e^{2\nu\,\left(x-d_1\nu\,t\right)}\medskip\\
+C_1e^{\nu\,\left(x-d_1\nu\,t\right)}+C_1^3\left(\nu\,\left(x-d_1\nu\,t\right)-3\right)e^{3\nu\,\left(x-d_1\nu\,t\right)}+
2\left(1+C_1\,e^{\nu\,\left(x-d_1\nu\,t\right)}\right)^3\ln\left(1+C_1\,e^{\nu\,\left(x-d_1\nu\,t\right)}\right)\Big]
, \ea\] where $\alpha$ and $C_1$ are arbitrary constants and
$\nu=\sqrt{\frac{-A_1}{2}}, \ \sigma = \frac{d_2 -d_1}{2 d_2}\,\nu.$

 Finally, making a simple analysis, we
conclude that the above exact solution cannot be obtained via
ans\"atze (\ref{3-2}) and (\ref{3-4}) provided $d_1\not=d_2$. Thus,
this exact solution  is not obtainable via the Lie method, i.e. is a
non-Lie solution (see  an extensive discussion on this matter in
Chapter 4 of \cite{ch-se-pl-book}).

\section{Proof of Theorem \ref{th-1} for the RD system (\ref{2-2}) }\label{sec-4}

In this section, we present the proof of Theorem \ref{th-1}, which
is the main theoretical result of this work. We start from
definition of $Q$-conditional (non-classical) symmetries.
 Let us
consider the most general form of two-component evolution systems
\begin{equation}\label{4-1}\ba  S_1 \equiv
u_{t} - F^1(u, v, u_x, v_x, u_{xx}, v_{xx})= 0, \\
S_2 \equiv v_{t} - F^2(u, v, u_x, v_x, u_{xx}, v_{xx})= 0,
\ea\end{equation} where $F^1$ and $F^2$ are smooth functions of all
arguments. In what follows, we assume that both of the above
equations are of second-order.


The general form of a $Q$-conditional symmetry is given by the
following expression
\[
Q= \xi^{0}(t,x,u,v) \partial_t + \xi^{1}(t,x,u,v) \partial_x +
\eta^{1}(t,x,u,v) \partial_u +  \eta^{2}(t,x,u,v) \partial_v.
\]
The relevant invariant surface conditions have the form
\[Q(u)\equiv \xi^{0}u_t + \xi^{1}u_x -
\eta^{1}=0, \  Q(v)\equiv \xi^{0}v_t + \xi^{1}v_x - \eta^{2}=0. \]

\begin{definition}\label{def} \cite{bluman2010,ch-dav-book}
The operator $Q$ is called a $Q$-conditional symmetry for an
evolution system if the following criterion is satisfied:
\begin{equation}\label{4-3}
Q^{(2)}(S_i)\Big|_M =0 ,\quad i=1,2.
\end{equation}
where the manifold $M$ is given by
\[\ba
M= \Big\{ S_i =0,i=1,2, \ Q(u)=0,\ Q(v)=0,\ \frac{\partial}{\partial
t} Q(u)=0,\ \frac{\partial}{\partial x} Q(u)=0,\\
\frac{\partial}{\partial t} Q(v)=0,\ \frac{\partial}{\partial x}
Q(v)=0 \Big\}, \ea\]
 and $Q^{(2)}$ is the second prolongation of the operator $Q$
\[
Q^{(2)}= Q^{(1)}  + \sigma^{1}_{tt}
\partial_{u_{tt}}+\sigma^{1}_{xx} \partial_{u_{xx}} +\sigma^{1}_{tx} \partial_{u_{tx}}
+\sigma^{2}_{tt} \partial_{v_{tt}}+\sigma^{2}_{xx} \partial_{v_{xx}}
+\sigma^{2}_{tx} \partial_{v_{tx}}+\sigma^{2}_{tt}
\partial_{v_{tt}}.
\]
while the first prolongation is
\[
Q^{(1)}= Q  + \rho^{1}_t \partial_{u_t} + \rho^{2}_t \partial_{v_t}
+ \rho^{1}_x \partial_{u_x} + \rho^{2}_x \partial_{v_x}.
\]
\end{definition}

\begin{remark} The classical definition of Lie symmetry is obtained
from the above criteria by replacing the manifold $M$ by $M_{Lie}=
\{ S_i =0,i=1,2\}$, i.e. making the manifold  much larger.
\end{remark}

The coefficients
 $\rho^1_t$, $\rho^2_t$, $\rho^1_x$,  $\rho^2_x$  and
 $\sigma^{1}_{tt}$, $\sigma^{1}_{xt}$, $\sigma^{1}_{xx}$,  $\sigma^{2}_{tt}$, $\sigma^{2}_{xt}, \sigma^{2}_{xx}$
 in the above formulae are calculated by the well-known expressions using the functions
  $\xi^{0}, \  \xi^{1}, \ \eta^{1}$
 and $\eta^2$ (see e.g.
 \cite{bluman2010,ch-se-pl-book,olv-93}).
   Actually, they were derived in the classical works by
 Sophus Lie.

 It is well-known that two different cases  occur
depending on the function $\xi^0(t,x,u,v)$:
\newline
Case 1. $\xi^0 \neq 0$;
\newline
Case 2. $\xi^0 =0,\quad \xi^1 \neq 0$.
\newline
These two cases need to be examined separately. It was shown earlier
\cite{ch-dav-book} that the manifold $M$ in Case 1 can be simplified
by omitting differential consequences of the invariant surface
conditions, i.e. the manifold
\begin{equation}\label{4-4*}
M^1 =\Big\{ S_1=0, \quad S_2=0,\quad  Q(u)=0, \quad Q(v) =0 \Big\}
\end{equation}
can be applied in the criteria (\ref{4-3}). It will be shown below
that relevant determining equations are completely integrable in the
case of the RD system (\ref{2-2}) and, as a result, Theorem
\ref{th-1} is obtained.

 Case 2, a so-called no-go case, is much more complicated for
analysis because  the determining equations obtained cannot be
solved in general (of course, some particular solutions are
obtainable). Thus, one may look for $Q$-conditional symmetries of
the first type \cite{ch-2010} instead of general $Q$-conditional
(non-classical) symmetries. Further analysis of Case 2 lies beyond
the scope of this study.

 Let us consider \textbf{Case 1}. Without losing  a generality,
  the operator $Q$  can be rewritten in  the following form
\begin{equation}\label{4-4**}
Q= \partial_t +  \xi(t,x,u,v) \partial_x + \eta^{1}(t,x,u,v)
\partial_u +
 \eta^{2}(t,x,u,v) \partial_v,
\end{equation} i.e.  $ \xi^0=1$.

In the general case, application Definition \ref{def} to systems of
the form  (\ref{4-1}) involving arbitrary functions leads to very
complicated systems of DEs, which typically are not integrable. It
should be noted that  the system of DEs is derived and briefly
discussed in  \cite{ch-dav-book} for  the standard RD system
\[\ba
u_{t}= d_1u_{xx} +F^{1}(u, v), \\
v_{t}= d_2v_{xx}+F^{2}(u, v). \ea\]




We now take a specialisation for the functions $F^1(u,v)$ and
$F^2(u,v)$ as follows:
\[\ba
F^1= d_1u_{xx}- A_1 u^2(1-u) - B_1 u v + 2 B_1 u^2 v + B_1 u v^2, \\
 F^2= d_2v_{xx} - B_2 v^2(1-v) - B_2 u v + 2 B_2 u v^2 + A_2 u^2 v,
\ea\] where the right-hand-sides correspond to the RD system
(\ref{2-2}).  Using Definition \ref{def} with the manifold from
(\ref{4-4*}) and making rather cumbersome calculation, it was shown
that the coefficients of the operator $Q$ (\ref{4-4**}) possess the
form
\[\ba
 \xi(t,x,u,v)= s(t,x), \\
 \eta^1(t,x,u,v)= p^1(t,x) +r^1(t,x) u + q^1(t,x) v, \\
 \eta^2(t,x,u,v)=p^2(t,x) +q^2(t,x) u + r^2(t,x) v,
\ea\] where the functions arising in the left-hand side must be
determined from the two sets of equations listed below.

The determining equations corresponding to the  first equation from
(\ref{2-2}) are
\begin{align}
&\frac{\left(d_2-d_1\right) q^1 p^2}{d_2}-d_1 p^1{}_{xx}+2 p^1 s{}_x+p^1{}_t=0,  \label{11} \\
&B_1 p^1-d_1 q^1{}_{xx}+\frac{\left(d_2-d_1\right) q^1 r^2}{d_2}+2 q^1 s{}_x+q^1{}_t=0,\label{12} \\
&-\frac{B_2 d_1 q^1}{d_2}+B_1 q^1-B_1 p^1=0, \label{13} \\
&\frac{B_2 d_1 q^1}{d_2}-B_1 q^1=0, \label{14}  \\
&2 A_1 p^1+B_1 p^2+\frac{\left(d_2-d_1\right) q^1 q^2}{d_2}-d_1 r^1{}_{xx}+2 r^1 s{}_x+r^1{}_t=0, \label{15}\\
&2 A_1 q^1-\frac{B_2 d_1 q^1}{d_2}+B_1 r^2-4 B_1 p^1-2 B_1 p^2+2 B_1 s{}_x=0, \label{16} \\
&\frac{2 B_2 d_1 q^1}{d_2}-4 B_1 q^1-2 B_1 r^2-2 B_1 s{}_x=0, \label{17} \\
&A_1 r^1-3 A_1 p^1+2 A_1 s{}_x+B_1 q^2-2 B_1 p^2=0, \label{18} \\
&\frac{A_2 d_1 q^1}{d_2}-3 A_1 q^1-2 B_1 r^2-2 B_1 r^1-2 B_1 q^2-4 B_1 s{}_x=0, \label{19} \\
&-2 A_1 r^1-2 A_1 s{}_x-2 B_1 q^2=0, \label{110} \\
&\frac{d_1 q^1 s}{d_2}-2 d_1 q^1{}_x-q^1 s=0, \label{111}\\
&-2 d_1 r^1{}_x+d_1 s{}_{xx}-s{}_t-2 s s{}_x=0,\label{112}
\end{align}
while those  corresponding to the  second equation are
\begin{align}
&\frac{\left(d_1-d_2\right) q^2 p^1}{d_1}-d_2 p^2{}_{xx}+2 p^2 s{}_x+p^2{}_t=0, \label{21} \\
&B_2 p^1+2 B_2 p^2+\frac{\left(d_1-d_2\right) q^1 q^2}{d_1}-d_2 r^2{}_{xx}+2 r^2 s{}_x+r^2{}_t=0,\label{22} \\
&B_2 \left(q^1+r^2-2 p^1-3 p^2+2 s{}_x\right)=0,\label{23} \\
&-2 B_2 \left(q^1+r^2+s{}_x\right)=0, \label{24} \\
&B_2 p^2-d_2 q^2{}_{xx}+\frac{\left(d_1-d_2\right) r^1 q^2}{d_1}+2 q^2 s{}_x+q^2{}_t=0, \label{25} \\
&-2 A_2 p^1-\frac{B_1 d_2 q^2}{d_1}+B_2 \left(r^1+2 q^2-4 p^2+2 s{}_x\right)=0, \label{26} \\
&-2 A_2 q^1+\frac{B_1 d_2 q^2}{d_1}-B_2 \left(2 r^2+2 r^1+3 q^2+4 s{}_x\right)=0, \label{27} \\
&q^2 \left(B_2-\frac{A_1 d_2}{d_1}\right)-A_2 p^2=0,\label{28} \\
&2 \left(\frac{B_1 d_2}{d_1}-2 B_2\right) q^2-2 A_2 \left(r^1+s{}_x\right)=0, \label{29} \\
&\left(\frac{A_1 d_2}{d_1}-A_2\right) q^2=0, \label{210} \\
&d_2 \left(s{}_{xx}-2 r^2{}_x\right)-s{}_t-2 s s{}_x=0, \label{211} \\
&\frac{\left(d_2-d_1\right) q^2 s}{d_1}-2 d_2 q^2{}_x =0.
\label{212}
\end{align}

It turns out that the systems  of equations (\ref{11})--(\ref{112})
and (\ref{21})--(\ref{212}) are solvable entirely, enabling the
result presented in Theorem~\ref{th-1} to be proved.

 The determining system of equations (that comes as
two sets of equations coming from the action of the second
prolongation on the two equations) is nonlinear, which is an
important distinction from the case of Lie symmetries.
 The structure of the partitioned sets of vanishing and non vanishing
 parameters plays an important role in finding all the subcases, to present a complete solution. First we distinguish the following two cases (I) and (II) depending on
parameters $d_1$ and $d_2$. It is chosen that both parameters do not
vanish as we are
 looking for non-trivial systems of evolutionary equations.
\newline
(I): $d_1=d_2$,
\newline
(II): $d_1 \neq d_2$.

For both cases (I) and (II), we then distinguish 5 cases; for which
all other parameters ($A_1$,$B_1$,$A_2$,$B_2$)  are non vanishing or
if one of them vanishes. The analysis is
 facilitated by constraints such as $A_1$ and $B_1$ not being simultaneously 0 and similarly
  for $A_2$ and $B_2$:\\
(1) $A_1,B_1,A_2,B_2 \neq 0,$
\newline
(2) $A_1=0$, $B_1 \neq 0$, $B_2$ or $A_2$ can be 0,
\newline
(3) $B_1=0$, $A_1 \neq 0$, $B_2$ or $A_2$ can be 0,
\newline
(4) $A_2=0$, $B_2 \neq 0$, $B_1$ or $A_1$ can be 0,
\newline
(5) $B_2=0$, $A_2 \neq 0$, $B_1$ or $A_1$ can be 0.\\
 This will
exhaust all possible cases (with some cases being the same or cases
being included in other ones more general). We denote simply the
cases as (1I), (2I), (3I), (4I), (5I) and  (1II), (2II), (3II),
(4II), (5II) respectively. We will then consider (a), (b), (c),
etc... for the branches and (1), ... for further branches. Here we
present details for the two main  cases leading to the result
presented in Theorem~\ref{th-1}. All other cases were examined in a
very similar way and it was shown that there are no other
$Q$-conditional symmetries.

\subsection{Case (3Ia)}

We have the constraints of parameters given by $B_1 =0$, $A_1 \neq
0$. Then we obtain directly that (\ref{110}) leads to $-2A_1 (r^1 +
s_x)=0$, which then provides $r^1 = -s_x$. The constraint given by
(\ref{111}) $-2d q^1_x =0$  implies that $q^1$ is only a function of
time, denoted as $q^1=q^1_0(t)$. The constraint (\ref{18}) $(-3 p^1
+s_x)=0$ yields $p^1 = \frac{1}{3} s_x$. Among the remaining
equations there are the constraints (\ref{29}) and (\ref{17})
\[ B_2 q^2_0(t)=0 ,\quad B_2 q^1_0(t) =0. \]
Taking $B_2=0$ and $A_2 \neq 0$ then we obtain from (\ref{26}) and
(\ref{27}) (similarly (\ref{16}))
\[ -\frac{2}{3} A_2 s_x =0 ,\quad -2 A_2 q^1_0(t) =0. \]
This implies $q^1_0(t)=0$ and $s=s_0(t)$. We then have as
consequence of (\ref{112}) that $s_0=l_1$ a constant. In addition
(\ref{212}) and (\ref{25}) give $-2dq^2_x=0$ and $q^2_t=0$ leads to
$q^2_0=l_2$ a constant. The equation (\ref{211}) leads to $-2d
r^2_x=0$ and $r^2=r^2_0(t)$. Finally, it can be shown from
(\ref{22}) that $r^2_0=l_3$, also a constant. Similarly, from
(\ref{28}) the function $p^2$ takes the form $p^2 =l_4$, where $l_4$
is a constant. The remaining equations ((\ref{21}), (\ref{28}) and
(\ref{210})) lead to
\[ -A_1 l_2 -A_2 l_4 =0 ,\quad (A_1 -A_2) l_2 =0, \]
which provide among solutions
\[ A_1= A_2 ,\quad l_4 = l_2 \]
and finally
\[ \xi= l_2,\quad \eta^1=0,\quad \eta^2 =-l_2 + l_2 u + l_3 v, \]
which then can be recast as
\[ Q= \partial_t + \gamma \partial_x+  \alpha (u-1) \partial_v + \beta v \partial_v.   \]
As it was noted above (see (\ref{2-5*})), the above symmetry  is a
Lie symmetry because $d_1=d_2$.

 It was also checked that $B_2 \neq 0$ does not lead to non-trivial symmetries.

\subsection{Case (3IIa)}
In this case $B_1=0$ and $A_1 \neq 0$, we have from (\ref{110})
$-2A_1(r^1+ s_x)=0$ which gives $r^1 =-s_x$. Another constraint,
equation (\ref{18}), gives $A_1 (r^1-3p^1+2 s_x)=0$
 then implies $p^1= \frac{s_x}{3}$. Then from (\ref{14}), (\ref{24}) and (\ref{23}) we obtain the following equations to be satisfied
\[ \frac{2 B_2 d_1 q^1}{d_2}=0 ,\quad -2 B_2 (q^1 + r^2 +s_x) =0 ,
\quad B_2 \left(q^1 + r^2 -3 p^2 + \frac{4}{3} s_x \right) =0. \] If
$B_2=0$ (the case $B_2\not=0$ again gives nothing) then  (\ref{27})
and (\ref{16}) lead to
\[ q^2 =-\frac{A_2 d_1 p^2}{A_1 d_2} ,\quad q^1 =0 .\]
We then obtain from (\ref{26}) $-\frac{2}{3} A_2 s_x =0$ and as
consequence $s=s_0(t)$. Then it can be shown from (\ref{112}) that
$s_0=l_1$. The equation (\ref{211})  reduces to $-2 d_2 r^2_x=0$ and
leads to $r^2=r^2_0(t)$. It can be demonstrated with (\ref{22}) that
$r^2=l_2$. Among remaining equations there is (\ref{212}) that leads
to
\[ (d_1-d_2) l_1 p^2 +2 d_1 d_2 p^2_x =0. \]
It can be solved as
\[ p^2 = p^2_0(t)\exp\Big( \frac{l_1 x}{2 d_1} - \frac{l_1 x}{2 d_2} \Big) . \]
Taking into account another remaining equation (\ref{110})
\[ A_2 ( A_2 d_1 - A_1 d_2) p^2 =0. \]
We also need for non-trivial function $p^2$ to set
\[ A_2= \frac{ A_1 d_2}{d_1}. \]
Finally we obtain from (\ref{21}) and (\ref{25}) the ODE
\[ -(d_1-d_2)^2 l_1^2 p^2_0(t) +4 d_1^2 d_2 (p^2_0)' =0. \]
The solution is given
\[ p^2_0(t)= \exp\left(\frac{(d_1-d_2)^2 l_1^2}{4 d_1^2 d_2}\,t\right) m_1. \]
Then taking $ d_1 = \frac{d_2 \gamma}{\gamma+ d_2 \sigma}$, $l_1 =
\gamma$, $m_1 =-\alpha$
 and $l_2 =\beta$ provide the result
\[ Q= \partial_t + \gamma \partial_x+   \Big( \alpha \exp( \sigma (x + d_2 \sigma t)) (u-1)
+ \beta v \Big) \partial_v .  \] Obviously, the above operator
coincides with that in (\ref{2-5}). The corresponding functions
$F^1$ and $F^2$ take the form
\begin{equation}\label{5-1*} \ba
F^1=d_1u_{xx}-A_1u^2(1-u), \\
F^2=d_2v_{xx}+A_1\frac{d_2}{d_1}u^2v.\ea \end{equation} Thus,
setting $A_1=Ad_1$, being $A$ an arbitrary constant, one derives
exactly the RD system (\ref{2-4}) presented in Theorem~\ref{th-1}.

This completes the proof of Theorem~\ref{th-1}.

\section{Conclusion}\label{sec-6}
 The main part of this study is devoted to the
RD system related to change in frequency of alleles in a population.
When mobility does not depend on genotype, despite some other
desirable attributes differing among the six possible genotypes
resulting
 from three possible alleles at a diploid site, the full system of RD equations for genotype populations reduces to a system
  of two cubic reaction-diffusion equations for two independent allele frequencies. To some extent, differences in mobility may
  then be accounted for in such a system with two different values of diffusion coefficient. In this case, one obtains
   the RD system  (\ref{2-2}), which is our main object of investigation.  A full classification of
  $Q$-conditional (non-classical) symmetry reductions has been completed, excepting the no-go case.
  Particular attention has been paid to the case of Mendelian
   inheritance with successive new alleles being dominant, therefore two special cases of the RD system
   (\ref{2-2}) were examined in detail. This has resulted in new reductions, both classical and non-classical,
   and in interesting new exact solutions involving a range of special functions such as elliptic integrals, the Lambert W function,
   and a general solution of the linear diffusion equation with (and without) a linear adsorption term.
    Some of the newly exposed special symmetric
    cases depart from the original motivating application. However, they do invoke possible new applications.  One of these is a  colonising human population
    that extracts a mineral that is mobile by erosion. Another is a two-species system with commensalism. For the application to population genetics, a more accurate
description would require more than two possible values of the
diffusion coefficient. This would require much more intensive
analysis, perhaps by stochastic simulation.

\section{Acknowledgments} IM was supported by Australian
Research Council Future Fellowship FT180100099. PB gratefully
acknowledges support from the Australian Research Council through
Discovery
 Project DP220101680. R.Ch. acknowledges that this research was funded by the British
Academy's Researchers at Risk Fellowships Programme and  by the
Ukrainian Visitor Program of the Sydney Mathematical Research
Institute (SMRI). V.D. acknowledges that this research was supported
by a grant from the Simons Foundation (SFI-PD-Ukraine-00014586,
V.D.).


\begin{thebibliography}{99}

\bibitem {aris-75I} \textsc{R. Aris},
 \textit{The Mathematical Theory of Diffusion and Reaction in Permeable Catalysts: the Theory of the Steady
 State},
 Clarendon Press, Oxford, 1975.

 \bibitem{a-h-b} \textsc{D.~J. Arrigo, J.~M. Hill  and P. Broadbridge},
Nonclassical symmetries reductions of the linear diffusion equation
with a nonlinear source, \textit{IMA J. Appl. Math.} \textbf{52}
(1994), 1--24.

\bibitem{bluman2010} \textsc{G.~W. Bluman, A.~F. Cheviakov  and S.~C. Anco},  \textit{Applications of Symmetry
Methods to Partial Differential Equations}, Springer,  New York,
2010.

\bibitem {britton}   \textsc{N.~F. Britton},  \textit{Essential Mathematical
bBiology},  Springer, Berlin, 2003.

\bibitem{h-b} \textsc{B.~H Bradshaw-Hajek and P. Broadbridge},  A robust cubic  reaction-diffusion system for gene
propagation,
 \textit{Math. Comp. Mod.} \textbf{39} (2004), 1151--1163.

\bibitem{broadbridge08}
\textsc{B.~H. Bradshaw-Hajek, P. Broadbridge and G.~H. Williams},
 Evolving gene frequencies in a population with three possible
alleles, \textit{Math. Comp. Mod.} \textbf{47} (2008), 210--217.

\bibitem{broadbridge07}
 \textsc{B.~H. Bradshaw-Hajek,  M.~P. Edwards, P. Broadbridge and G.~H. Williams},
Nonclassical symmetry solutions for reaction-diffusion equations
with explicit spatial dependence, \textit{Nonlinear Anal.}
\textbf{67} (2007), 2541--2552.

\bibitem{BHH} \textsc{P. Broadbridge, B.~H. Bradshaw-Hajek and A.~J. Hutchinson},  Conditionally integrable PDEs,
 non-classical symmetries and applications, \textit{Proc. R. Soc. A} \textbf{479} (2023), 20230209.

\bibitem{broadbridge23}
\textsc{P. Broadbridge,  R. Cherniha and J. Goard},  Exact
nonclassical symmetry solutions of Lotka--Volterra type population
systems, \textit{Euro. J. Appl. Math.} \textbf{34} (2023),
998--1016.

\bibitem{ch-2010} \textsc{R. Cherniha},   Conditional symmetries
 for systems of PDEs:  new definition and their application for
 reaction-diffusion systems,  \textit{J. Phys. A Math.
 Theor.} \textbf{43} (2010), 405207.

\bibitem{ch-dav-book}\textsc{R. Cherniha and  V. Davydovych},
\textit{Nonlinear Reaction-Diffusion Systems --- Conditional
Symmetry, Exact Solutions and Their Applications in Biology},
Lecture Notes in Mathematics  \textbf{2196}, Springer, Cham,  2017.

\bibitem{cherniha21b}
\textsc{R. Cherniha and  V. Davydovych},  New conditional symmetries
and exact solutions of the diffusive two-component Lotka--Volterra
system,
 \textit{Mathematics} \textbf{9} (2021), 1984.

\bibitem{ch-da-EJAM-22} \textsc{R. Cherniha and  V. Davydovych},  A reaction–diffusion system with
cross-diffusion: Lie symmetry, exact solutions and their
applications in the pandemic modelling, \textit{Euro. J. Appl.
Math.} \textbf{33} (2022), 785--802.

\bibitem{ch-da-AAM-23} \textsc{R. Cherniha and  V. Davydovych},  Symmetries and exact solutions of
the diffusive
 Holling--Tanner prey-predator model, \textit{Acta Appl. Math.}
 \textbf{187} (2023), 8.

\bibitem {2-ch-king1}  \textsc{R. Cherniha and  J.~R. King},
 Lie   symmetries of nonlinear  multidimensional
reaction-diffusion systems: I, \textit{J. Phys. A: Math. Gen.}
\textbf{33} (2000), 267--282.

\bibitem {2-ch-king2} \textsc{R. Cherniha and  J.~R. King},
 Lie   symmetries of nonlinear  multidimensional
reaction-diffusion systems: II, \textit{J. Phys. A: Math. Gen.}
\textbf{36} (2003), 405--425.


\bibitem {ch-ki-24} \textsc{R. Cherniha and  J.~R. King},  Nonlinear systems of PDEs admitting
infinite-dimensional Lie algebras and their connection with Ricci
flows, \textit{Stud Appl Math.} \textbf{153} (2024), e12737.

\bibitem {ch-ki-25} \textsc{R. Cherniha and  J.~R. King},  Nonlinear systems of PDEs admitting
infinite-dimensional Lie algebras and their connection with Ricci
flows. II: The two-dimensional space case, \textit{Stud Appl Math.}
\textbf{155} (2025) , e70120

\bibitem {ch-se-pl-book} \textsc{R. Cherniha,  M.  Serov and O. Pliukhin},
 \textit{Nonlinear Reaction-Diffusion-Convection Equations: Lie and
Conditional Symmetry, Exact Solutions and Their Applications},
Chapman and Hall/CRC, New York, 2018.


\bibitem{cla-mans-94}  \textsc{P.~A. Clarkson and  E.~L. Mansfield},
   Symmetry reductions and exact solutions of a class of
nonlinear heat equations, \textit{Phys. D} \textbf{70} (1994),
250--288.

\bibitem {fife-79} \textsc{P.Fife},   \textit{Mathematical Aspects of Reacting and Diffusing
Systems},
 Springer, New York, 1975.

 \bibitem {oliveri-et-al-25} \textsc{M. Gorgone , F. Oliveri and E.
Sgroi}, Hierarchy of coupled Burgers-like equations induced by
conditional symmetries, \textit{Z. Angew. Math. Phys.}  \textbf{76}
(2025), 8.

\bibitem {ku-na-ei-16} \textsc{Y. Kuang, J.~D. Nagy and  S.~E. Eikenberry},  \textit{Introduction to Mathematical
Oncology},
 CRC Press, Boca Raton, 2016.

 \bibitem{lo-dimas-bo-2023} \textsc{E. Lopez, S. Dimas and Y. Bozhkov},  Symmetries of Ricci
 flows,
 \textit{Adv. Nonlinear Anal.} \textbf{12} (2023), 20230106.

\bibitem {mur2}  \textsc{J.~D. Murray},  \textit{Mathematical Biology},
Springer, Berlin, 1989.

\bibitem {mur2003} \textsc{J.~D. Murray}, (2003)
\textit{Mathematical Biology II}, Springer, Berlin, 2003.


\bibitem{naz-24} \textsc{R. Naz, A.~G. Johnpillai  and F.~M. Mahomed},  The exact solutions of a diffusive SIR model via symmetry
groups,
 \textit{J. Math.}  \textbf{2024} (2024), 4598831.

\bibitem {okubo} \textsc{A. Okubo and S.~A. Levin},  \textit{Diffusion and Ecological Problems.
Modern Perspectives, 2nd edn}, Berlin, Springer, 2001,


\bibitem {olv-93}   \textsc{P. Olver},  \textit{Applications of Lie
Groups to Differential Equations},  Springer, Berlin, 1993.


 \bibitem{Pat-Wint} \textsc{J. Patera and  P. Winternitz}, Subalgebras of real three-and
four-dimensional Lie algebras, \textit{J. Math. Phys.} \textbf{18}
(1977), 1449--1455.

 \bibitem{rosa-23} \textsc{M. Rosa, M.~L. Gandarias, A. Ni\~{n}o-L\'{o}pez and S. Chuli\'{a}n},
   Exact solutions through symmetry reductions for a
high-grade brain tumor model with response to hypoxia, \textit{Chaos
Solitons Fractals} \textbf{171} (2023),  113468.

\bibitem{soph-24} \textsc{C. Sophocleous},  Non-Lie reduction operators and potential
transformations for a special system with applications in plasma
physics, \textit{Symmetry} \textbf{16} (2024), 207.

\bibitem{torrisi-21} \textsc{M. Torrisi and  R. Tracina},  Lie
symmetries and solutions of reaction-diffusion systems arising in
biomathematics, \textit{Symmetry} \textbf{13} (2021), 1530.

\bibitem{torrisi-23} \textsc{M. Torrisi and  R. Tracina},  Symmetries and solutions
for a class of advective reaction-diffusion systems with a special
reaction term, \textit{Mathematics} \textbf{11} (2023), 160.



\bibitem {zhdanov}  \textsc{R.~Z. Zhdanov and V.~I. Lahno},  Conditional symmetry
of a porous medium equation,   \textit{Phys. D} \textbf{122} (1998),
178--86.

\end{thebibliography}
\end{document}